**Highlights**

- First systematic review on GenAI and LLMs in entrepreneurship research
- Identify five thematic clusters using TF-IDF, PCA, and hierarchical clustering
- Discuss ethical challenges related to GenAI in Entrepreneurship
- GenAI and LLMs act as external enablers by transforming entrepreneurial activity
- Outline future research on GenAI in Entrepreneurship

# GenAI in Entrepreneurship

## a systematic review of generative artificial intelligence in entrepreneurship research: current issues and future directions

Anna Kusetogullari [1,*], Huseyin Kusetogullari [2], Martin Andersson [3], Tony Gorschek [1]

[1]Department of Software Engineering, Blekinge Institute of Technology, Karlskrona, Sweden

[2]Department of Computer Science, Blekinge Institute of Technology, Karlskrona, Sweden

[3]Department of Industrial Economics, Blekinge Institute of Technology, Sweden and Swedish Entrepreneurship Forum, Stockholm

***Corresponding author**: Anna Kusetogullari (**E-mail**: anna.kusetogullari@bth.se)

# GenAI in Entrepreneurship

## a systematic review of generative artificial intelligence in entrepreneurship research: current issues and future directions

## Abstract

Generative Artificial Intelligence (GenAI) and Large Language Models (LLMs) are recognized to have significant effects on industry and business dynamics, not least because of its impact on the preconditions for entrepreneurship. There is yet a lack of knowledge of GenAI as a theme in entrepreneurship research. This paper presents a systematic literature review aimed at identifying and analysing the evolving landscape of research on the effects of GenAI on entrepreneurship. We analyse 83 peer-reviewed articles obtained from leading academic databases: Web of Science and Scopus. Using natural language processing and unsupervised machine learning techniques with TF-IDF vectorization, Principal Component Analysis (PCA), and hierarchical clustering, five major thematic clusters are identified: (1) Digital Transformation & Behavioural Models, (2) GenAI-Enhanced Education & Learning Systems, (3) Sustainable Innovation & Strategic AI Impact, (4) Business Models & Market Trends, and (5) Data-Driven Technological Trends in Entrepreneurship. Based on the review, we discuss future research directions, gaps in the current literature as well as ethical concerns raised in the literature. We pinpoint the need for more "macro-level" research on GenAI and LLMs as external enablers for entrepreneurship and research on effective regulatory frameworks that facilitate business experimentation, innovation and further technology development.

# 1. Introduction

Generative AI (GenAI) and large language models (LLMs), such as GPT-4, are often claimed to have a potential to re-shape the preconditions for business dynamics and entrepreneurship by automating a broader set of tasks, enhancing decision-making, enabling rapid innovation and boost efficiency (Rezazadeh et al 2025, Uriarte et al 2025, Liu and Wang 2024).[1] For both existing and aspiring entrepreneurs, the technologies can lower entry barriers, reduce startup costs, and accelerate the development cycle of new ventures (Dwivedi et al. 2021). It is fair to say to that GenAI and LLMs fall in the category of 'external enablers' for entrepreneurship (see e.g. Kimjeon and Davidsson, 2022).

Against this backdrop one can expect that the potential effects of GenAI and LLMs are high on the agenda in entrepreneurship research. Understanding how technologies like GenAI and LLMs influence the potential for entrepreneurship, how they are used by entrepreneurs, how they influence entrepreneurial opportunities and business strategies as well as what ethical and strategic challenges they present are essential issues for both entrepreneurship scholars and practitioners. However, despite a rapid expansion, the academic contribution literature on GenAI and entrepreneurship remains fragmented. There are several papers studying the applications of AI in business and entrepreneurship, but rather few papers focused on GenAI (Giuggioli & Pellegrini, 2023; Chalmers et al., 2020; Gupta et al.,2023). While the benefits of GenAI in terms of efficiency, new business models and innovations are widely discussed, issues related to ethics, digital maturity, and the contextual variability of AI integration have received less attention.

This paper makes five key contributions to the literature. First, we present a comprehensive systematic literature review focused on the intersection of GenAI and entrepreneurship. Second, the study introduces a novel, unsupervised machine learning-based approach, combining Term Frequency–Inverse Document Frequency (TF-IDF) vectorization, Principal Component Analysis (PCA), and hierarchical clustering to thematically classify the literature. This approach is applied to analyze 83 peer-reviewed articles retrieved from Web of Science and Scopus. Third, we categorize the literature into five thematic domains that define the current intellectual landscape: Digital Transformation & Entrepreneurial Behaviour, GenAI-Enabled Education & Learning Systems, Strategic AI for Sustainable Innovation, Evolving Business Models & Market Trends, Data-Driven Technologies in Entrepreneurial Contexts. Fourth, we assess key trends, discuss the ethical issues related to usage of GenAI in entrepreneurship, and emerging research gaps. Finally, we outline future research directions for researchers and entrepreneurs.

---

[1] GenAI can for example be used to generate business plans, conduct market analysis, and even prototype software or marketing materials.

# 2. Background

This article explores two widely used models in the context of entrepreneurship: Generative AI (GenAI) and Large Language Models (LLMs).

## 2.1 Generative Artificial Intelligence (GenAI): A brief overview

GenAI is a subset of Artificial Intelligence that creates new data based on patterns learned from existing datasets (Jebara, 2012). Unlike traditional models that primarily focus on detection (Pinto and Sobreiro. 2022), forecasting (Schade et al., 2023), and decision-making (Maine et al., 2015), generative models are designed to create new instances that resemble the original data. This ability makes them highly valuable in various entrepreneurial applications, including product design, content creation, customer service automation, business process optimization, and market analysis (Korzynski, 2023; Corvello, 2024).

Deep generative models, i.e. a class of learning algorithms designed for data generation, have significantly transformed numerous domains in engineering and the sciences, enabling advances in fields ranging from cybersecurity to materials discovery (Yinka-Banjo & Ugot, 2020; Regenwetter et al., 2022; Bilodeau et al., 2022; Guo & Zhao, 2022; Liu et al., 2023; Yan & Li, 2023).

For entrepreneurs, GenAI presents transformative opportunities by enabling rapid prototyping of ideas, reduced operational costs, and a potential for enhanced decision-making (Korzynski , 2023). Startups and businesses can leverage AI-generated content to develop marketing campaigns, generate product descriptions, automate customer interactions, and create engaging media assets. AI-powered chatbots and virtual assistants help improve customer service, ensuring 24/7 support and personalized responses based on historical data and user preferences.

A distinguishing feature of GenAI is its capability for unsupervised learning, which allows it to recognize and replicate patterns without requiring labelled datasets. This is especially beneficial when acquiring labelled data is difficult or expensive. For instance, GenAI can generate synthetic market research data, simulating consumer behaviour to provide insights for product development and business strategies. Additionally, in financial modelling, GenAI assists entrepreneurs in predicting trends, assessing risks, and optimizing investment strategies.

Beyond business efficiency, GenAI plays a crucial role in innovation-driven entrepreneurship. It can be used in developing new business models by identifying gaps in the market and generating novel solutions. For aspiring entrepreneurs, understanding and integrating GenAI into their ventures provides a competitive edge. Whether through automating repetitive tasks, enhancing creativity, or generating data-driven insights, AI-driven solutions enable businesses to scale efficiently and remain adaptable in an increasingly AI-powered market landscape. Embracing these technologies can lead to more innovative, agile, and resilient business models, shaping the future of entrepreneurship (Mariani and Dwivedi, 2024).

## 2.2 Large Language Models: A brief overview

LLMs are a subset of GenAI and have emerged as highly capable systems, demonstrating remarkable proficiency in various Natural Language Processing (NLP) tasks (Corvello, 2024). The introduction of advanced models such as OpenAI's GPT-4, Gemini, DeepSeek, Anthropic's Claude, and Google's PaLM2 have significantly accelerated advancements in both NLP and the broader field of artificial intelligence (Maarouf et al., 2025; Lang et al., 2024). There is a speculation that these models are approaching human-level performance. LLMs like GPT-4 operate on an autoregressive framework, which enables them to generate sequences such as sentences by predicting each subsequent word based on prior context. Unlike traditional language models, LLMs benefit from more sophisticated training techniques and enhanced capabilities (Chang et al., 2024; Zhao et al. 2025). They rely on Transformer-based architectures, which use an 'attention' mechanism to assess the significance of words within a sentence. This allows them to evaluate the entire context of a given text, making them exceptionally effective for sequential predictions.

The LLMs are trained on vast amounts of textual data, learning to predict words by continuously refining internal parameters to minimize the gap between their predictions and actual language patterns (Chang et al., 2024; Zhao et al. 2025). A significant advantage of LLMs is their ability to perform various language tasks without requiring additional specialized training. Since they have been pre-trained on extensive datasets, they can generate meaningful and contextually appropriate responses based solely on user input.

One of their practical applications is in text summarization, where LLMs can analyse and condense large bodies of text into concise summaries (Chang et al., 2024; Zhao et al. 2025). These summaries can serve as high-quality benchmarks for evaluating other summarization models. Beyond summarization, LLMs are employed in tasks such as text classification, which automatically sorts text into predefined categories. This capability is useful in areas such as spam detection and sentiment analysis, where the model can efficiently process and categorize vast amounts of data. Another application involves evaluating authorship and attribution. LLMs can assist in determining the origin of a text, whether identifying the author of an anonymous document or verifying whether a social media post was written by a particular individual (Zhao et al. 2025).

## 2.3 Summary

In summary, one can argue that GenAI and LLMs are examples of external enablers for entrepreneurship. External Enablers in entrepreneurship refers to environmental factors outside the entrepreneur's control that can enable or facilitate the emergence, development, and success of entrepreneurial ventures (Kimjeon and Davidsson 2022).

GenAI and LLMs influence entrepreneurship in at least two ways: (i) 'doing current work more effectively', for example by reducing costs or time spent on given tasks, and (ii) 'doing new things'. The latter refers to that these technologies can create new opportunities and pave the way for new business models, new products and services, and transform the market environment (cf. Rezazadeh et al 2025, Uriarte et al 2025).

# 3. Methodology

## 3.1 Materials and Research Design

To conduct a comprehensive analysis of the literature on GenAI and entrepreneurship, we employ a systematic literature review (Tranfield et al., 2003).

The process of formulating our search query followed an iterative approach, centring on two key categories of terms: one associated with "entrepreneurship" and another with "generative artificial intelligence." The keyword deployed to cover entrepreneurship is "entrepreneur*" or "startup" or "start-up". The list of keywords used to cover generative artificial intelligence included: "Generative Artificial Intelligence" or "Generative AI" or "GPT" or "GPT-1" or "GPT-2" or "GPT-3" or "GPT-4" or "ChatGPT" or "gemini" or "deepseek" or "GrammarlyGO" or "BlackBox" or "DALL-E 2" or "Synthesia" or "DeepBrainAI" or "Runway" or "Bard" or "Cohere Generate" or "Claude" or "StyleGAN" or "Bardeen" or "Rephrase.ai" or "Descript" or "Type Studio" or "GLIDE" or "Imagen" or "Bidirectional Encoder Representations from Transformers" or "T5" or "BERT" or "RoBERTa" or "ERNIE" or "Bart" or "large language models" or "Generative Pre-trained Transformer" or "Llama" or "LLM" or "chatbot" or "Generative Adversarial Networks" or "Diffusion Models" or "Mixtral" or "BigGAN" and various other GenAI technical methods.

Our review focused exclusively on peer-reviewed journal articles published in English up until March 06, 2025. As shown in Figure 1, we retrieved studies from two major academic databases, Scopus and Web of Science (WOS), using the specified search terms within article titles, abstracts, and keywords.

The initial search yielded a total of 688 research papers including articles, review articles, proceeding papers and book chapters: 511 from Scopus and 177 from WOS. In the next phase, we narrowed our search by only considering articles and review papers in English language that covered the subject areas of "Business" and "Management" for WOS and "Business, management and accounting", "Economics, econometrics and finance" and "Decision sciences" for Scopus. As a result, we yielded a total of 158 articles: 101 from Scopus and 57 from WOS. Next, we identified and eliminated duplicate records, ensuring that articles appearing in multiple databases were included only once in our final dataset. This refinement reduced the total number of articles to 110. We then conducted an abstract screening to assess their relevance. During this stage, we identified that 27 articles did not focus on the intersection of GenAI and entrepreneurship. As a result, the final selection comprised 83 articles.

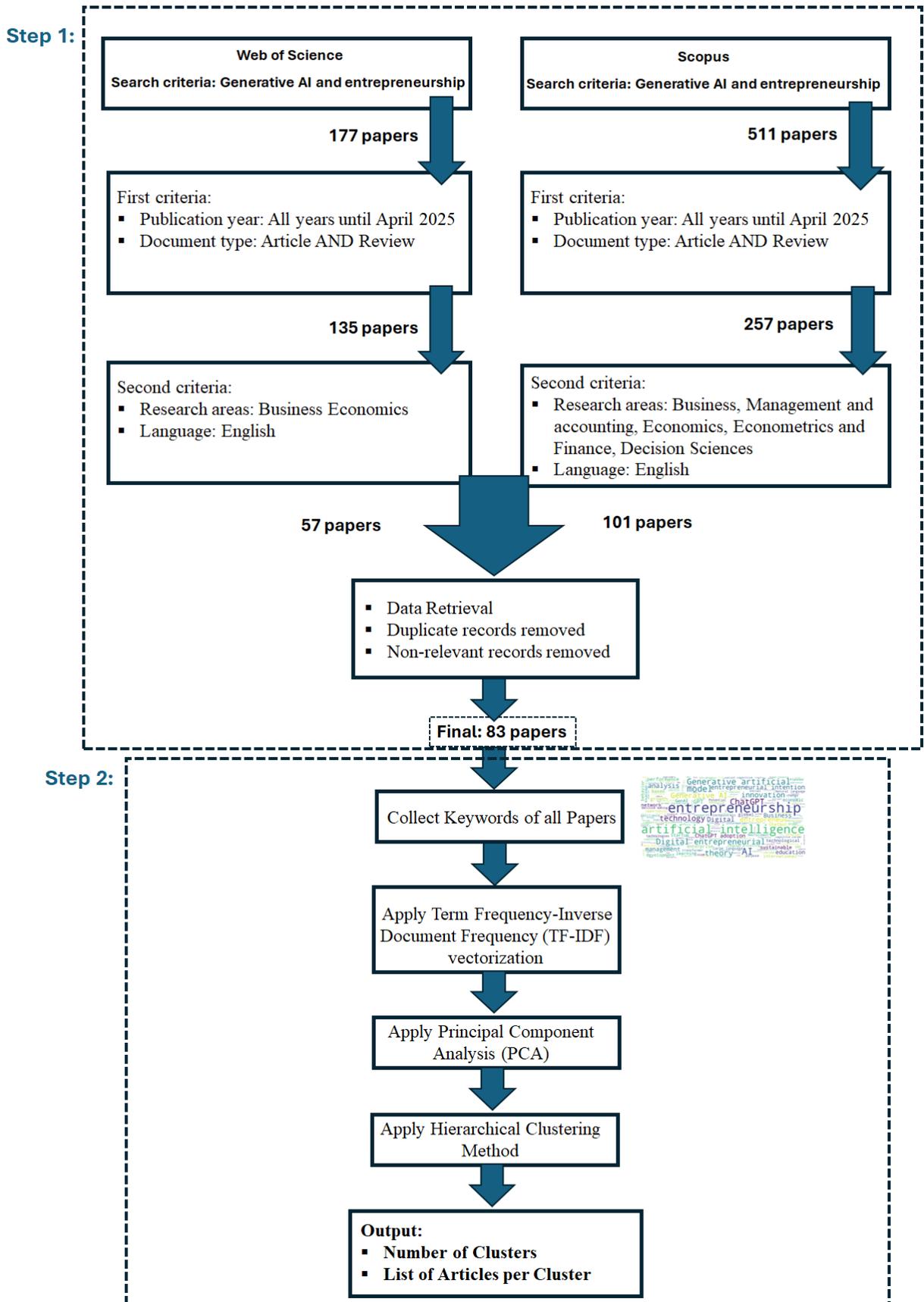

**Figure 1:** Procedure for paper search, collection and clustering research articles.

## 3.2 Clustering Approach

To conduct analysis and clustering on the filtered articles related to GenAI and entrepreneurship, we utilized a combination of advanced natural language processing and unsupervised machine learning techniques. A key component of our methodology involved transforming the textual content into a numerical representation using term frequency–inverse document frequency (TF-IDF) vectorization (Leskovec et al., 2019). This step helps emphasize the relative importance of specific terms within the dataset, forming a solid foundation for subsequent analytical processes.

After that, we use Principal Component Analysis (PCA) (Lever et al., 2017) to reduce the dimensionality of the resulting TF-IDF vectors. This reduction preserves the essential structure of the data while removing noise which makes the clustering process more efficient and interpretable. Then, we employ hierarchical clustering technique (Patel et al., 2015) to group the articles based on their keyword lists. Unlike K-means clustering method (Sinaga and Yang, 2020), which requires a predefined number of clusters, hierarchical clustering allows for greater flexibility in exploring the data's natural groupings without a prior predefined number of clusters. Additionally, it provides a dendrogram, offering visual insight into the nested relationships among topics. Compared to the DBSCAN clustering method (Ester et al. 1996), which is sensitive to parameter settings and may struggle with varying cluster densities, hierarchical clustering proves more stable and better suited for high-dimensional data when combined with PCA. As a result, Hierarchical clustering in combination with Principal PCA is well-suited for our study. Besides this, this technique is widely recognized and frequently applied in various domains, including anomaly detection, pattern recognition, and others (Stefato and Hamza, 2007; Jafarzadegan et al., 2019; Granato et al., 2018). Note that the clustering approach is conducted using Python.

The algorithm starts by treating each paper as its own cluster and then iteratively merges the most similar clusters based on similarity, building a hierarchical tree structure. This process continues until all papers are grouped under a single root node. The clustering results are visualized using a dendrogram, where each leaf node represents a paper, and each branch indicates a merge step as shown in Figure 2. The y-axis denotes the distance or dissimilarity between papers or clusters, while the x-axis lists the individual papers, labeled according to their assigned cluster themes. Based on this approach, the papers are grouped into five different clusters as shown with different colors of clusters in the dendrogram.

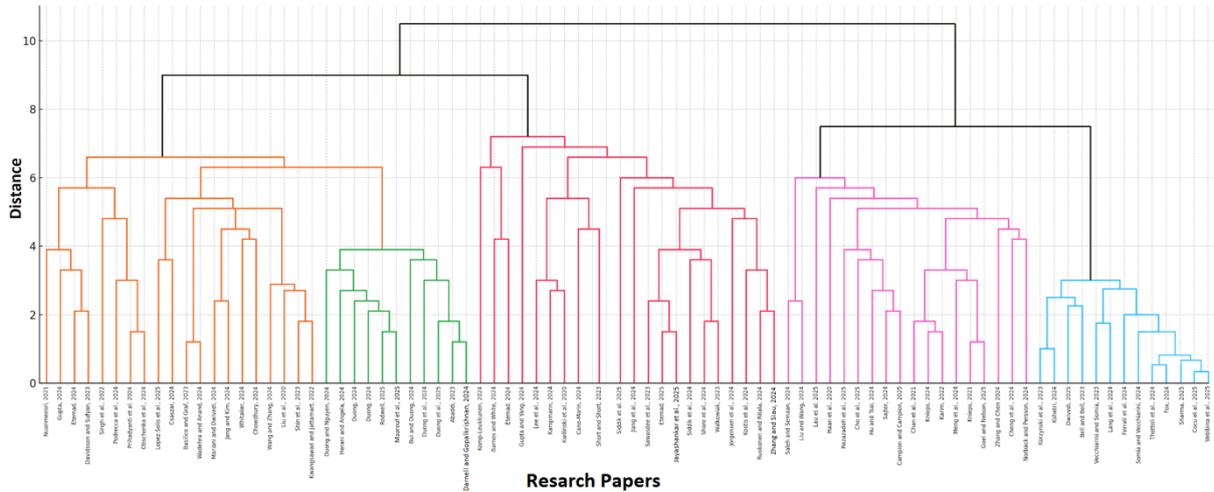

**Figure 2:** Dendrogram Showing the Distribution of 83 Research Papers Using PCA-Based Hierarchical Clustering.

Figure 2 presents a dendrogram resulting from integration of PCA and hierarchical clustering performed on 83 research papers. The visualization depicts five different clusters, each representing a group of similar studies, as identified through patterns of shared keywords. The x-axis displays individual papers labeled by reference, while the y-axis represents the degree of dissimilarity, with lower vertical positions indicating greater similarity between papers or clusters.

The green cluster, labeled "Digital Transformation & Behavioral Models," includes papers such as Duong and Nguyen (2024) and Herani and Angela (2024), which merge at a low height, suggesting a high degree of thematic overlap. For instance, both papers explore the intersection of ChatGPT adoption and digital entrepreneurship among young individuals in emerging economies. As the hierarchy progresses, closely related papers are grouped into intermediate subclusters, which merge into broader thematic categories. Moreover, the blue cluster, "Generative AI-Enhanced Education & Learning Systems," reflects shared focus on GenAI-driven educational innovations.

Higher-level mergers illustrate convergence among broader domains. For instance, the red cluster, "Business Models & Market Trends," reflects research on enhancing business models, formulating business strategies, and analysing market trends. This cluster highlights how GenAI and LLMs are reshaping competitive positioning and entrepreneurial engagement within entrepreneurial ecosystems. Meanwhile, the orange cluster, "Sustainable Innovation & Strategic AI Impact," brings together research on green entrepreneurship and strategic applications of GenAI. The purple cluster, "Data-Driven Tech Trends in Entrepreneurship," reflects intersections between technological innovation and advanced data analytics, capturing how entrepreneurs use big data, GenAI, and other emerging technologies to inform decision-making, optimize operations, and scale digital ventures.

# 4. Findings

## 4.1 Descriptive Statistics

The clustering analysis of 83 research articles on GenAI in Entrepreneurship using Hierarchical and PCA provides critical insights into how research in this domain is structured and evolving. By clustering papers based on their keywords, five distinct research themes emerged:

(1) Digital Transformation & Behavioural Models
(2) GenAI-Enhanced Education & Learning Systems
(3) Sustainable Innovation & Strategic AI Impact
(4) Business Models & Market Trends
(5) Data-Driven Tech Trends in Entrepreneurship.

The distribution of papers across these clusters reveals both the dominant areas of research and potential gaps that may require further academic attention.

### 4.1.1 Distribution of Articles Published by Cluster Topic

Figure 3 depicts the distribution of 83 research papers across five distinct clusters derived from their keyword profiles using PCA and hierarchical clustering technique.

The largest cluster, Business Models & Market Trends, includes 21 papers. This cluster highlights the entrepreneurial applications of GenAI, emphasizing how businesses leverage AI-driven innovations to reshape traditional business practices and market strategies. Key topics include AI-powered decision-making, emerging business model innovations, investor influences, and AI-driven entrepreneurial communication. The considerable size of this cluster suggests growing scholarly interest in the transformative potential of AI technologies on entrepreneurship and market dynamics.

The second-largest cluster is Sustainable Innovation & Strategic AI Impact, comprising 20 articles. The publications in this vein reflects a significant academic focus on the role of GenAI in driving sustainable business practices, innovation strategies, and long-term competitive advantage. Research in this cluster mostly explores how GenAI facilitates innovation management, enhances organizational sustainability, and addresses strategic considerations within entrepreneurial ecosystems.

Data-Driven Tech Trends in Entrepreneurship, consisting of 18 papers, represents another research area. This cluster examines advanced technological applications, such as ChatGPT, to enhance entrepreneurial decision-making processes and outcomes. It particularly emphasizes LLM's role in forecasting business performance, assessing entrepreneurial risks, and improving crowdfunding strategies, reflecting an increased reliance on AI-driven methodologies in entrepreneurial research.

The GenAI in Education & Learning Systems cluster, comprising 13 papers, explores how GenAI tools like ChatGPT are integrated into entrepreneurial education, curriculum development, and competency enhancement. This cluster indicates a rising trend toward using AI to facilitate innovative and experiential learning methods that bolster entrepreneurial skills such as creativity, critical thinking, and decision-making. Lastly, the Digital Transformation & Behavioural Models cluster, with 11 papers, investigates entrepreneurs' behavioral responses and technology adoption strategies associated with GenAI. Themes include digital self-efficacy, behavioural intention theories, and technostress, particularly regarding the adoption and integration of AI technologies within entrepreneurial ventures. This cluster underscores ongoing research into how behavioural factors influence the effective implementation and utilization of GenAI in business contexts.

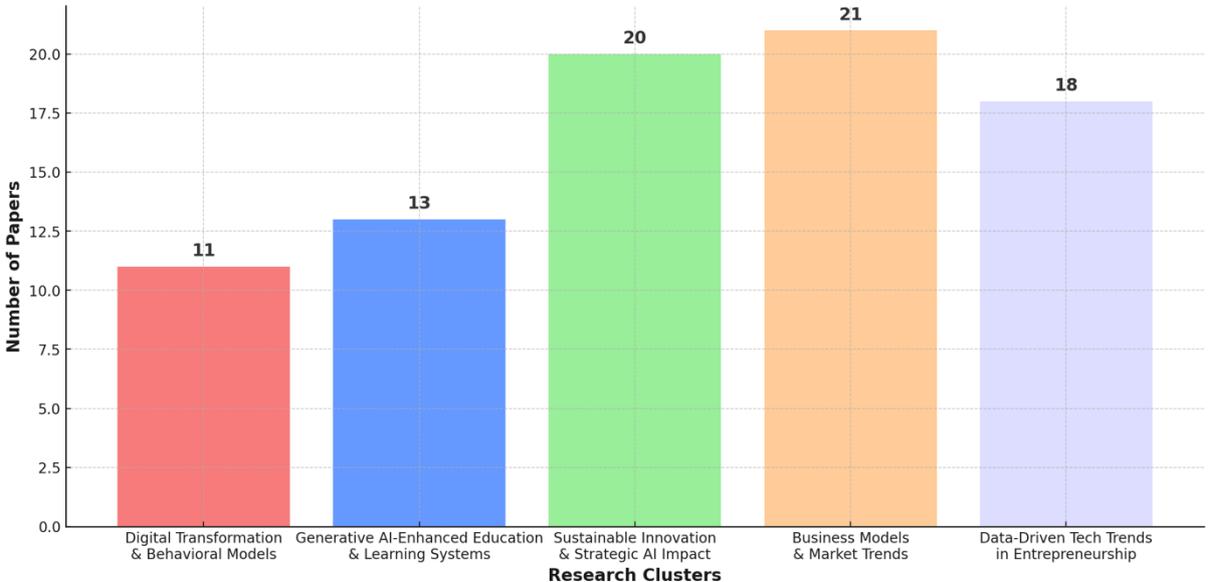

**Figure 3:** Distribution of research articles across five different clusters.

### 4.1.2 Distribution of Articles Published by Year

Figure 4 demonstrates the yearly distribution of research articles related to GenAI and LLMs in entrepreneurship from 2020 to 2025. Between 2020 and 2022, research activity using GenAI in entrepreneurship was relatively limited, reflecting the early stages of GenAI adoption and its emergence as a mainstream tool. Only three research articles were published in each of the years 2020 and 2021, indicating limited activity during this initial period. In 2023, 11 articles were published and this growth increased in 2024, reaching a peak of 47 articles, indicating a big academic engagement with GenAI's role in entrepreneurship.

Furthermore, during the first few months of 2025, 17 articles have been published suggesting a continuing growth, potentially surpassing 2024 by year-end if the current pace continues. It highlights the rapidly growing interest among entrepreneurs in using GenAI and LLM methods to shape entrepreneurial strategy, drive innovation, support decision-making, anticipate market trends, enhance education and others.

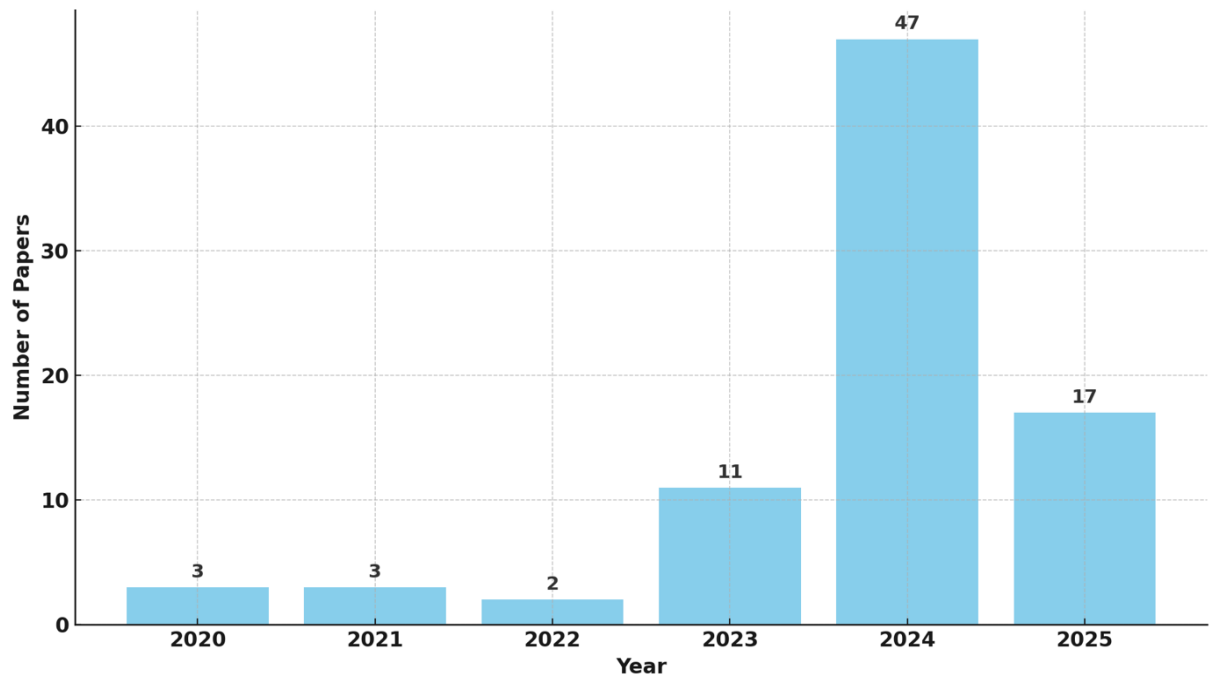

**Figure 4:** Number of articles published per year.

## 4.2 Cluster 1: Digital Transformation & Behavioural Models

Advanced AI models such as GenAI and ChatGPT, are reshaping the entrepreneurial landscape by transforming not only the mechanics of business creation but also the cognitive and behavioural foundations of entrepreneurship. Current research in this area provides diverse insights into how digital technologies intersect with entrepreneurial behaviour, emphasizing how psychological, emotional, and social factors mediate the use of GenAI tools in entrepreneurial contexts.

The adoption of GenAI remains a challenging task for entrepreneurs. The stream of papers in this cluster study different issues of adoption and focus on entrepreneurship behaviour. Herani and Angela (2024) investigate adoption among Indonesian youth, revealing that performance expectancy and perceived usefulness are key motivators, while effort expectancy serves as a significant barrier. Notably, their study uncovers a paradox: simultaneous perceptions of high effort and high benefit increase motivation, highlighting the complex behavioural dynamics behind digital entrepreneurship in emerging markets.

Several studies explore how ChatGPT influences entrepreneurial cognition and intention. Duong and Nguyen (2024) demonstrate that ChatGPT enhances opportunity recognition, entrepreneurial knowledge, and self-efficacy, all of which contribute to stronger entrepreneurial intentions. Extending this behavioural approach, Bui and Duong (2024) apply Social Cognitive Career Theory to show that entrepreneurial self-efficacy mediates the link between ChatGPT usage and intention, while technostress weakens this relationship. This underscores the emotional and cognitive burdens that entrepreneurs face in AI-driven environments.

Meanwhile, Duong (2024b) introduces the concept of digital entrepreneurial identity aspiration, showing how ChatGPT fosters identity formation, which mediates entrepreneurial action. However, this process is impeded by technology-related anxiety, illustrating the psychological barriers to AI adoption.

Further refining the behavioural framework, the Theory of Planned Behaviour (TPB) is employed to investigate the psychological and social mechanisms behind ChatGPT adoption. Using data from 604 Vietnamese university students, Duong et al. (2024) find that ChatGPT positively shapes attitudes, subjective norms, and perceived behavioural control, which mediate the relationship between ChatGPT adoption and digital entrepreneurial intention. These findings reflect how psychological pathways can support effective technological engagement. Rodwell (2025) reinforces this view, arguing that behavioural factors such as emotional responses, user goals, and cultural communication patterns must be embedded in GenAI design to create tools that are intuitive, trustworthy, and inclusive.

A further contribution integrates the Technology Acceptance Model with TPB to examine how GenAI tools affect entrepreneurial intention. Using data from 400 business students and structural equation modelling, the study finds that both perceived usefulness and behavioural control predict intention, reinforcing the value of combining technological and psychological models for a holistic view of AI adoption (Abaddi, 2023). Another study by Duong et al. (2024) expands this framework by showing that ChatGPT adoption enhances self-efficacy, which then boosts positive attitudes and reinforces entrepreneurial commitment—capturing the full behavioural cycle from AI interaction to entrepreneurial action. Maarouf et al. (2025) develop a fused LLM using BERT to predict startup success by combining structured venture data with unstructured textual descriptions from VC platforms. The model captures behavioural and communicative information embedded in startup narratives. The study illustrates how AI-driven text analytics are transforming investment decision-making and behavioural modeling in startup ecosystems, shifting the process from intuition-based judgment to data-informed strategies.

This literature cluster illustrates that GenAI is not merely a tool for automation, but a transformative force tied to human cognition, emotion, and behaviour. Embedding behavioural models within AI adoption frameworks is therefore essential for enabling inclusive, and adaptive resonant entrepreneurial ecosystems.

## 4.3 Cluster 2: Generative AI-Enhanced Education & Learning Systems

The integration of GenAI, particularly tools such as ChatGPT or Gemini, into education and learning systems is changing not just the mode of delivery, but also the foundations of entrepreneurial education and research. The studies in this cluster reveal a converging narrative of opportunities, challenges, and institutional shifts driven by the emergence of GenAI.

One line of research examines the integration of GenAI in entrepreneurial education, underscoring the importance of institutional support for student learning and application of digital tools. GenAI is positioned as a strategic tool, assisting student entrepreneurs with different tasks, such as market research, forecasting, and business planning. Thottoli et al.

(2023) offer strategic recommendations and identify a gap in the application of GenAI tools in academic incubators.

At the instructional level, GenAI is also reshaping entrepreneurship classrooms—impacting both content delivery and teaching methodologies. There is a growing need for instructors to ensure that students use it responsibly. Bell and Bell (2023) stress the need to support future entrepreneurs in becoming capable, reflective users of GenAI tools.

GenAI enhances personalized learning, creativity, and self-efficacy for entrepreneurs, supporting tasks such as opportunity recognition, business model generation, storytelling, and decision-making. A systematic review by Dwivedi (2025) confirms the role of GenAI in entrepreneurship, particularly in education, innovation, and ethical integration. The review highlights GenAI's ability to facilitate iterative learning, foster entrepreneurial competencies, and enable strategic development. However, it also raises important concerns about overreliance, data ethics, and the need for critical, balanced engagement.

From a practical view, GenAI is rapidly reshaping how entrepreneurship is taught, making it essential for instructors to adapt their approaches. Darnell and Gopalkrishnan (2024) shares a hands-on experience with that process, offering a thoughtful approach to creating value for entrepreneurship students. It also offers a set of flexible entrepreneurship exercise templates that can be used across a wide range of courses to engage diverse learners (Darnell and Gopalkrishnan, 2024).

Some studies highlight the role of ChatGPT in entrepreneurship education and show that it benefits in enhancing creativity, supporting idea generation, and helping in business plan development (Vecchiarini and Somià, 2023; Somià and Vecchiarini, 2024; Sharma, 2025). However, there is a risk of overreliance, highlighting the need for critical engagement with AI outputs and active guidance from educators. For instance, Fox (2024) propose to document how AI tools like ChatGPT are used in studies, ensuring methodological rigor and supporting reproducibility in academic research.

Even the integration of LLMs marks a shift toward intelligent-assisted learning that is accessible anytime, the complex nature of entrepreneurship education poses challenges for AI systems. In particular, it needs improvement in tasks requiring semantic reasoning and advanced mathematical computation. Lang et al. (2024) highlight the need for algorithmic optimization and model enhancements to address these challenges.

To support effective integration of GenAI tools, Korzynski et al. (2023) introduce a theoretical framework for prompt engineering, which is a strategic skill for customizing and directing LLMs outputs. The framework underscores how students can become co-creators with AI by learning to customize and direct outputs effectively. In line with this, Cocu et al. (2025) propose a cloud-based educational model that incorporates GenAI-enabled tools, green innovation frameworks, and AI-assisted business planning platforms. This approach is designed to embed sustainability into engineering education, offering a scalable and future-oriented framework that equips students with the tools needed for eco-conscious entrepreneurship. Complementing this pedagogical innovation, Vebibina et al. (2025) evaluate ChatGPT's capability to generate complex entrepreneurship assessments.

From an institutional perspective, the shift toward GenAI adoption is analyzed through the lens of institutional theory. Key actors, referred to as institutional entrepreneurs, play a crucial role in reframing AI's place in employability and educational innovation (Kshetri, 2024). Another approach to using ChatGPT in academic research is presented through a "4D Framework"—Discover, Develop, Discuss, and Deliver, which encourages dynamic engagement with AI tools to augment scholarly activities (Ferrati et al., 2023).

Together, these studies illustrate how GenAI is advancing a more dynamic and competency-focused model of entrepreneurship education and creates a new trend in entrepreneurial education.

## 4.4 Cluster 3: Sustainable Innovation & Strategic AI Impact

The intersection of sustainability and GenAI is becoming an emerging literature area in entrepreneurship. Recent research illustrates how GenAI not only accelerates innovation but also enables green transitions, reshapes decision-making, and transforms the entrepreneurial process. Studies in this cluster highlight GenAI's potential as a technological catalyst and as a tool requiring careful human and institutional alignment for sustainable impact.

One research stream explores how GenAI tools interconnect with innovation frameworks to support entrepreneurial creativity and problem-solving. Shin et al. (2024) show how combining a structured innovation methodology with ChatGPT can enhance access to problem-solving processes and support sustainable decision-making in competitive environments. Focusing on elderly entrepreneurs, Nuanmeesri (2021) demonstrates how integrating GenAI chatbots, social CRM tools, and augmented reality into a digital marketing framework can dramatically enhance online business capabilities. With over 98% accuracy in customer engagement, the model demonstrates how AI can promote inclusive and sustainable economic participation.

Building on this framework, Etemad (2023) takes a global perspective, examining how international mall and medium-sized enterprises (SMEs) must adapt to increasing environmental, technological, and socio-economic disruptions. GenAI, along with technologies like IoT and Web 3.0, is shown to catalyse strategic redirection and business model innovation in response to climate change (Etemad, 2023). Another study reveals that integration of emerging technologies within regional knowledge systems contributes to decentralized innovation, especially as traditional sectors lose dominance to more diverse innovation clusters (Basilico and Graf, 2023).

Understanding technology adoption is another critical angle. Kwangsawad and Jattamart (2022) use a hybrid framework that blends the Innovation-Decision Process with the Technology Acceptance Model and identify key psychological barriers to innovation, such as privacy concerns and technological anxiety. These findings point to the importance of aligning AI tools with user values to ensure long-term adoption.

For the startup ecosystems, Gupta (2024) explores key adoption drivers. The study identifies factors such as domain expertise, system quality, and training that influence how GenAI tools are utilized. Incremental adoption is recommended, allowing startups to align experimentation with broader sustainability and value-creation goals. Chowdhury et al. (2024) propose to embed GenAI into workforce practices using institutional entrepreneurship theory, where GenAI is comparable to talent development, organizational learning, and sustainable innovation. This positions human capital and AI as interdependent forces in competitive advantage creation. However, there is a need for evolving strategies that can adapt to GenAI's dynamic contributions to entrepreneurial ecosystems (Davidsson and Sufyan, 2023).

GenAI can act as a change-agent for green innovation and knowledge management. Wang and Zhang (2024) show that GenAI improves green knowledge acquisition and innovation, with government support acting as a key moderator. Singh et al. (2022) demonstrate how firms use technologies such as chatbots and voicebots to drive sustainable, scalable innovation, enhancing service delivery while promoting low-carbon, high-impact business models. This points to GenAI's potential in advancing sustainability goals in green ventures.

As GenAI technologies mature, traditional industries are increasingly exploring their integration to drive sustainable innovation. Legal firms, for instance, are undergoing transformation, as evidenced by a recent study on Vidhi Partners and their deployment of proprietary GenAI tools, VidAI and VidAI Pro. This case illustrates how legacy organizations can embed AI into their core operations while maintaining employee trust and customer alignment. The study outlines a balanced approach to innovation that can serve as a practical framework for digital transformation within conventionally structured institutions (Wadehra and Anand, 2024). In the domain of talent acquisition, Jang and Kim (2024) examine the trade-offs between GenAI-powered recruitment platforms and traditional hiring practices. The findings suggest that a hybrid model combining both AI capabilities with human judgment can not only enhance operational efficiency but also reduce systemic bias, especially in small firms with limited resources.

Another critical dimension of GenAI adoption lies in strategic decision-making. While GenAI excels at pattern recognition and predictive analysis, a study by López-Solís et al. (2025) underscores the continued importance of human oversight, particularly in uncertain or high-stakes business contexts. The research advocates for a complementary model, wherein AI augments rather than replaces human insight, ensuring more resilient and ethically sound outcomes. Extending these insights, Csaszar et al. (2024) reveal that GenAI tools can match human entrepreneurs in generating and evaluating business strategies. However, the research cautions that long-term success still hinges on adaptability and contextual responsiveness—traits that remain uniquely human.

Together, these studies reflect a growing consensus: while GenAI offers powerful tools for efficiency and innovation, its most sustainable applications arise from thoughtful integration with human values, expertise, and adaptability. Furthermore, Obschonka et al. (2024) argue that long-term sustainable innovation will depend not only on technological advances but also on ethical implementation and inclusive research design. Delving into future research agendas, Mariani and Dwivedi (2024) identify ten core themes at the intersection of GenAI and innovation management. These include product development, regulatory strategies, IP protection, and ecosystem design, offering a foundation for advancing sustainable competitive strategies using GenAI.

Recent literature also emphasizes the use of GenAI in enabling sustainable innovation across sectors. For instance, Podrecca et al., 2024 analyse 5,919 patents to identify six major GenAI-related domains: predictive analytics, material sorting, defect detection, advanced robotics, scheduling, and resource optimization. Their findings point to GenAI's central role in climate-conscious manufacturing and energy efficiency. In agriculture, agri-tech start-ups are emerging as catalysts for social innovation, enabling sustainable growth through technology diffusion in rural economies (Prihadyanti et al., 2024). From a governance perspective, Whitaker (2024) highlights the need for sustainable, inclusive ownership models in GenAI ecosystems, advocating for hybrid structures that balance innovation with the public good. Together, these studies demonstrate that sustainable innovation with GenAI thrives at the intersection of strategic design and collaborative ecosystems.

## 4.5 Cluster 4: Business Models & Market Trends

The evolution of business models and market dynamics in entrepreneurship is increasingly shaped by digitalization, AI integration, and policy landscapes. This stream of studies illustrate how new models are emerging, how markets are responding, and how entrepreneurs adapt to rapidly changing contexts.

One critical area of focus is the reconfiguration of business models among SMEs navigating turbulent environments. Empirical study of French SMEs illustrates that GenAI adoption fosters entrepreneurial resilience and agility. However, the relationship between entrepreneurial orientation and resilience is negatively moderated by market turbulence, pointing to a non-linear interaction between strategic intent and market dynamics (Shore et al., 2024). This underscores the necessity for context-sensitive AI strategies in business model design.

Complementing this, a microeconomic model developed by Walkowiak (2023) explores how task interdependencies between humans and AI affect workplace productivity. The study categorizes firms into "open learning organizations" that integrate GenAI collaboratively and "closed learning organizations" that resist such integration. This typology offers valuable insight into how firms' structural and cultural orientations influence their capacity to leverage GenAI for business innovation.

Digital asset forecasting and startup financing remains a challenging task. Siddik et al. (2024) conduct a large-scale funding analysis of over 500 GenAI startups, revealing that investor networks play a more critical role in determining financial success than technological capabilities alone. Their findings suggest that sustained growth for GenAI-driven ventures depends not only on innovation but also on the strategic integration of investor engagement within business models. In the context of digital economy, Kwilinski et al. (2020) demonstrate how classical finance models combined with cloud-based systems can enhance short-term cryptocurrency predictions.

GenAI is also redefining startup support systems. Research on Thai entrepreneurial ecosystems shows that chatbot technologies, when integrated into incubator programs, enhance access to knowledge and advisory services. The perceived usefulness and service quality of AI virtual assistants are found to be pivotal factors influencing adoption (Sawasdee et al., 2023). This points to a paradigm shift in how early-stage ventures access foundational support—moving from in-person mentorship to AI-driven systems. Moreover, entrepreneurs use ChatGPT as a support for iterative innovation and strategic thinking (Kostis et al., 2024). The communicative dimension of entrepreneurship is also being reshaped. Short and Short (2023) examine how tools like ChatGPT are enabling entrepreneurs to craft distinctive brand narratives through prompt engineering. By modifying rhetorical styles, entrepreneurs can differentiate their ventures in competitive markets, opening new pathways for strategic identity. GenAI can also be used as a tool to study firms' market and operational resilience (Jorgensen et al., 2024). Overall, it suggests that GenAI can serve as an educational and strategic bridge in business model development.

On the global markets, international SMEs face complex learning curves in integrating GenAI. Etemad (2025) argues that experiential learning and continuous capability development are essential for successful adoption. A related theoretical contribution by Etemad (2024) reconceptualizes GenAI not merely as a tool but as a strategic co-pilot—encouraging firms to set clear goals within available resources and experiment with digital tools to adapt their global business models dynamically.

New conceptualizations are expanding the understanding of hybrid business models shaped by GenAI. Zhang and Siau (2024) propose a framework for "meta-entrepreneurship" that combines GenAI with metaverse technologies, spanning infrastructure, content, and user experience. These models reflect the growing prevalence of immersive, hybrid ecosystems where digital and physical value chains converge.

Further emphasizing GenAI's role in advancing business models, Ruokonen and Ritala (2024) outline three strategic archetypes which are digital tycoons, niche carvers, and asset augmenters, each representing distinct pathways through which firms embed GenAI into operations to gain data leverage, algorithmic efficiency, and executional strength. These strategies signal a shift toward AI-centric value creation across diverse industries. At the macroeconomic level, Siddik et al. (2025) establish a strong link between GenAI financing and financial development in 21 countries, with notable impacts in Asia. Their findings suggest that investor networks and media visibility significantly enhance startup development and ecosystem expansion.

Meanwhile, Komp-Leukkunen (2024) reveals that GenAI integration is redefining entrepreneurial practices within digital startups. GenAI reduces skill barriers, accelerates minimum viable product development, and enables leaner operational structures, effectively broadening participation in software-driven innovation (Jayashankar et al., 2025). Together, these contributions demonstrate GenAI's growing influence on financial flows, strategic differentiation, and structural shifts in entrepreneurial markets.

Other research highlights GenAI's impact on brand dynamics, growth patterns, and venture capital narratives. Barnes and White (2024) investigate the implications of access-based business models, revealing that while subscriptions open new revenue avenues, they can unintentionally weaken brand equity among consumers with strong group-brand connections. This trade-off between expansion and perceived brand commitment is especially relevant as startups experiment with flexible, AI-enhanced offerings. Lee et al. (2024) categorize mobile startups into distinct growth patterns such as 'rapid scalers' and 'niche dominators', based on user behaviour, showing that business model types (e.g., platform vs. pay-per-use) play a pivotal role in shaping growth trajectories. These distinctions help clarify which models align best with GenAI-driven agility. From a critical finance perspective, Kampmann (2024) unpacks the venture capital ecosystem's growing reliance on AI as a speculative and symbolic asset. He illustrates how inflated valuations often emerge from techno-optimism rather than operational viability, especially when GenAI is packaged as intangible value.

Further extending the strategic implications of GenAI for innovation, Cano-Marin (2024) uses natural language processing and network analysis to evaluate GenAI's impact on innovation ecosystems through patent publication. The findings confirm that GenAI enhances decision-making and productivity but also highlight limitations related to data availability and linguistic diversity. In fact, another study shows how digital entrepreneurship and GenAI tools are expanding opportunities for female-led ventures (Jiang et al., 2024).

Taken together, these studies offer a view of how GenAI is reshaping entrepreneurial business models and market behaviour. They highlight both the enabling potential, and the strategic complexities associated with GenAI integration, suggesting that future entrepreneurship will increasingly depend on the ability to adapt, collaborate with AI systems, and engage with evolving digital and institutional landscapes.

### 4.6 Cluster 5: Data-Driven Tech Trends in Entrepreneurship

The evolving landscape of entrepreneurship is increasingly shaped by data-driven technologies that transform decision-making, resource utilization, and innovation strategies. A range of recent studies reflects how entrepreneurs and firms are leveraging AI including GenAI and LLMs, natural language processing, and ICT infrastructure to build resilience, improve outcomes, and adapt to complex environments.

In entrepreneurial finance, LLMs offer diverse applications. Chan et al. (2021) demonstrate the practical potential of natural language processing by applying BERT models to predict crowdfunding outcomes. Their findings challenge the traditional emphasis on readability, showing that machine-assessed linguistic features are stronger predictors of funding success. This offers actionable guidance for entrepreneurs seeking to optimize communication and align messaging strategies with AI-driven performance indicators. Expanding on GenAI's role in decision-making, Saleh and Semaan (2024) apply sentiment analysis using LLMs such as ChatGPT and FinBERT to assess entrepreneurial sustainability. Findings indicate that positive sentiment, when accurately extracted from financial text, correlates strongly with firm survival, highlighting the predictive power of AI in evaluating entrepreneurial viability.

At a macroeconomic level, research emphasizes the importance of integrating multi-source data analytics to enhance economic resilience through financial inclusion. Hu and Tsai (2024) argue that combining real-time behavioural and transactional data offers a better understanding of entrepreneurial ecosystems, guiding policymakers in supporting entrepreneurship in the face of global uncertainties

A meta-analysis grounded in the Resource-Based View (RBV) explores the contextual performance of ICT and AI tools. Karim et al. (2022) find that while general-purpose and enabling technologies improve firm outcomes, their effectiveness depends on national digital maturity. This suggests that the entrepreneurial value of such tools is highly dependent on the quality of a country's digital infrastructure. Adding to the RBV perspective, Liu and Wang (2024) study university startups in China and find that tangible, intangible, and human resources related to GenAI substantially improve entrepreneurial performance. The study highlights internal integration and external collaboration as key mediators, reinforcing the view of GenAI as a strategic organizational asset when embedded effectively within the firm's resource ecosystem.

From a policy perspective, recent research explores how smart infrastructure shapes the foundation of digital innovation ecosystems. Knieps (2024) highlights that coordinated success to 5G environments acts as a critical enabler for decentralized, data-intensive entrepreneurial activity, especially in sectors like mobility and healthcare. Complementary work by Knieps (2021) further underscores the transformative impact of 5G and IoT technologies in facilitating real-time vertical and horizontal integration across industries. The low latency and high bandwidth of 5G networks are shown to enhance connectivity, agility, and innovation within entrepreneurial ecosystems in fields such as transport, energy, and logistics.

Moreover, GenAI tools are increasingly applied to service-based entrepreneurship. Cheng et al. (2024) introduce a sentiment analysis framework that combines BERT with the Kano model to extract actionable insights from customer feedback. While originally developed for the hospitality sector, this approach demonstrates how aspect-based sentiment analytics can help entrepreneurs refine service offerings and enhance customer experience through AI-driven feedback loops.

At the level of public awareness, research uses internet search trends to map entrepreneurial familiarity with AI. Results show that AI awareness is significantly higher in urban and affluent US regions, suggesting that geographic disparities influence AI adoption and entrepreneurial readiness. The distinction between general AI awareness and applied familiarity with tools like ChatGPT offers a foundation for assessing regional readiness (Goel and Nelson, 2025).

GenAI is also recognized as a driver of startup growth, facilitating product-led, sales-driven, and operational efficiency approaches through tools like LLMs, with use cases ranging from personalized content to customer journey optimization (Rezazadeh et al., 2025). In the context of SMEs, Cho et al. (2025) demonstrate that GenAI enhances digital resilience and management quality by aligning AI-powered tools such as automation and predictive analytics with operational challenges in volatile environments. A case study during the COVID-19 pandemic further shows how startups use data analytics and flexible budgeting to adapt business models, optimize cash flow, and navigate uncertainty (Zhang and Chen, 2024).

In more complex technical contexts, advanced GenAI architectures like GATv2 and ERNIE-GEN are shown to support integration in highly technical systems, expanding GenAI's role beyond standard business applications (Meng et al., 2024). Complementing this, Lau et al. (2025) propose a model that combines ontology-based AI, chatbots, and 3D data visualization in the metaverse. It supports entrepreneurs to manipulate data collaboratively and derive insights through immersive, AI-guided platforms.

Institutional models also reflect this trend. De Haan et al. (2020) analyse the Runway Startup Postdoc Program, showing how structured data-informed mentorship and translational research pathways within universities can accelerate deep-tech venture creation. At the regional level, a bibliometric analysis by Sajter, D. (2024) highlights a rising interest in entrepreneurship and quantitative economics, aligning with broader efforts to leverage AI and analytics for economic insight.

In a methodological review, automated text analysis techniques like LLMs are being adopted in entrepreneurship and HR research to enhance construct validity and enable scalable assessment of traits, identity, and team dynamics (Campion and Campion, 2025). Finally, a theoretical contribution by Norbäck and Persson (2024) proposes that while GenAI lower innovation costs, it may also increase barriers to entry due to data centralization. To maintain healthy innovation dynamics, the authors advocate for open-source ML initiatives that ensure broader entrepreneurial access, making creative destruction more equitable and sustainable (Norbäck and Persson, 2024).

Overall, studies included in this cluster reveal a dynamic shift toward data-driven entrepreneurship, where GenAI and advanced analytics are not only enhancing operational efficiency but also redefining how entrepreneurs access, interpret, and act on information. The landscape of entrepreneurial practice is becoming increasingly shaped by real-time data, algorithmic insight, and intelligent automation. These technologies empower founders to adapt quickly, collaborate more effectively, and scale innovations with precision—even in uncertain or resource-constrained environments.

However, this transformation also introduces critical challenges. Risks of data monopolization, ethical issues, and the need for methodological rigor call for deliberate policy, infrastructural, and open-source strategies to ensure inclusive, transparent and ethically aligned ecosystems that leverage the benefits of GenAI.

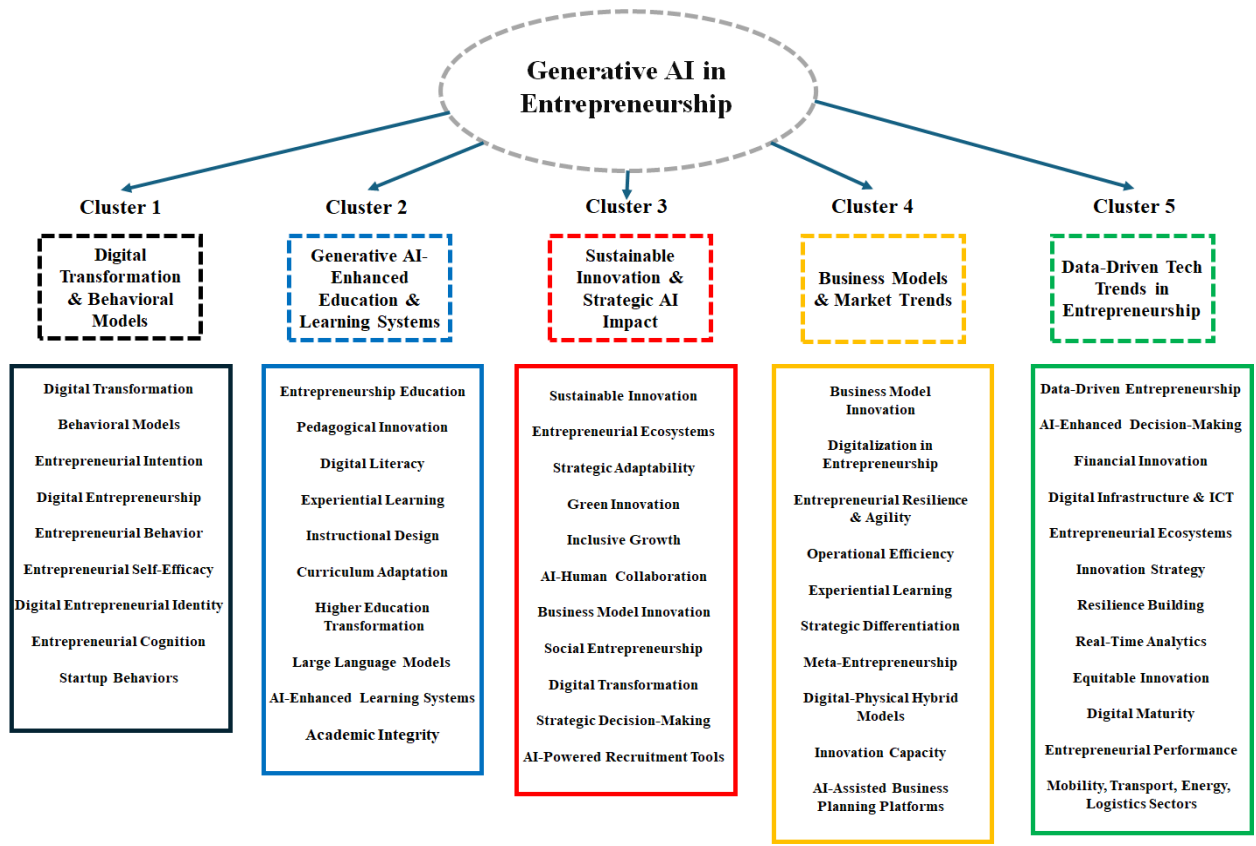

**Figure 5:** Common Keywords used in five different clusters in Generative AI Entrepreneurship Research.

# 5. Ethics, Opportunities and Future Directions

## 5.1 Ethics using Generative AI in Entrepreneurship

As discussed in the previous section, the integration of GenAI in entrepreneurial contexts provide potential benefits for innovation, product development, and automation. Besides, it helps entrepreneurs to improve operational efficiency, enhances marketing capabilities, and boosts overall competitiveness, giving firms a strategic advantage (Prasad Agrawal, 2023; Chui et al., 2022). However, alongside these benefits, the implementation of GenAI in tech-driven products and services raises important ethical concerns (Al Halbusi et al., 2023; Lin et al., 2020; Kumar et al., 2025; Maiti et al., 2025). Issue of ethics are related to questions about mitigating risks, promoting trust, upholding stakeholder interests, and supporting innovation, i.e. questions of relevance for aspiring as well as established entrepreneurs and organizations. .

One of the key ethical considerations in the use of GenAI and LLMs in entrepreneurship is data privacy and intellectual property protection (Chen et al. 2023, Han and Yin, 2025). As these models are often trained on vast datasets, including user-generated content and potentially proprietary information, it is crucial to ensure that innovative business ideas or sensitive data shared with such systems are not stored, reused, or exposed without consent. In entrepreneurial contexts, users may input novel concepts, strategies, or product designs into platforms like ChatGPT, expecting confidentiality. Therefore, a common argument is that developers and platform providers need data privacy strategies. Protecting the originality and confidentiality of entrepreneurial ideas is essential to maintaining trust and supporting innovation in this growing field.

Transparency in the usage of GenAI-produced outputs is crucial for raising and maintaining trust and acceptance among stakeholders. This involves entrepreneurs being transparent about objectives, data handling practices, and the limitations of GenAI technologies. Studies show that users are more likely to engage with GenAI-driven products when they trust the ethical orientation of the entrepreneur and perceive the firm as reliable and accountable (Przegalińska et al., 2019; Munoko et al., 2020). Entrepreneurs must strike a careful balance between openness and data privacy, particularly when dealing with sensitive information or potentially manipulative content such as deepfakes.

Another important ethical issue is the risk of plagiarism associated with GenAI-generated content. These models can produce highly human-like text, images, and code, making it easier to reproduce existing work without proper attribution (Hari et al., 2025; Vecchietti et al., 2025). Entrepreneurs must ensure originality in marketing materials, business proposals, and product development to avoid violating intellectual property rights and eroding creative integrity.

GenAI systems rely on large datasets for training, which may include biased or discriminatory content (Hari et al., 2025). If not carefully managed, these biases may be reproduced or even amplified by the GenAI, leading to ethical concerns around fairness and discrimination. Entrepreneurs using GenAI for decision-making such as in hiring, customer segmentation, or product personalization, must analyse and test outputs to prevent biased outcomes. It raises the risk of users' misinformation or damage brand reputation. A common argument is that there is a need for proactive approaches to recognize, mitigate, and eliminate these kinds of biases to ensure inclusivity, equity, and fairness in AI-driven solutions.

GenAI's ability to generate persuasive, realistic content can be exploited for unethical purposes such as manipulation, fake reviews, misleading advertisements (Vecchietti et al., 2025; Diro et al., 2025). Entrepreneurs must ensure that GenAI tools are not used to deceive consumers or misrepresent product offerings. Upholding authenticity in GenAI-driven communications is essential to maintaining ethical business practices and long-term consumer trust.

To address these and other challenges, regulatory frameworks are evolving. The European Commission's Ethics Guidelines for Trustworthy AI (2019) is based on the idea to explicate core principles including fairness, accountability, transparency, and respect for fundamental rights (Hari et al., 2025). However, guidelines and regulatory frameworks must be adapted to fit the entrepreneurial landscape and be designed in such ways that they do not obstruct technology development and business experimentation.

## 5.2 Future Research Directions: GenAI and LLMs in Entrepreneurship

This section outlines research directions on GenAI in Entrepreneurship across five thematic clusters identified and discussed in the systematic review. Each cluster highlights distinct domains, ranging from digital transformation and education to sustainability, business model innovation, and data-driven technologies. Table 1 presents key themes in the existing clusters, research gaps and/or limitations, future research directions and opportunities as well as potential methods within each area, offering a comprehensive agenda to advance research on GenAI in entrepreneurship.

**Table 1.** Future Research Directions on Generative AI in Entrepreneurship

| Cluster | Key Themes/Insights | Identified Gaps | Future Research Direction | Potential Methods / Theories |
|---|---|---|---|---|
| **1 Digital Transformation & Behavioral Models** | GenAI is reshaping entrepreneurial cognition, self-efficacy, and opportunity recognition processes, particularly within digitally transforming ventures. | Existing research employs cross-sectional survey methodologies, limiting insights into longitudinal cognitive and behavioral change. | Future studies should employ longitudinal and experimental research designs to examine how sustained interaction with GenAI influences entrepreneurial cognition, strategic behavior, and venture performance over time. | Longitudinal designs; Experimental methods; Social Cognitive Career Theory (SCCT); Theory of Planned Behavior (TPB) |
| | | Current participant samples are overly concentrated in early-stage entrepreneurs from emerging economies. | Future research should broaden geographical scopes by incorporating diverse entrepreneurial populations, including experienced founders, female entrepreneurs, and actors in developed economies. | Comparative cross-cultural designs; Contextual entrepreneurship frameworks |
| | | Conceptual framing is dominated by TAM and TPB, with limited interdisciplinary integration. | Scholars should consider integrating a broader set of theories such as Self-Determination Theory, Identity Theory, and Diffusion of Innovations to capture motivational, identity-based, and adoption-related mechanisms. | Interdisciplinary theory development; Qualitative and mixed-method designs |
| | | Emotional and psychological responses to GenAI remain underexplored. | Future research may examine how emotional states such as technostress, anxiety, and digital fatigue influence long-term GenAI adoption, resilience, and entrepreneurial persistence. | Psychology-informed models; Mixed-method approaches |

| | | Entrepreneur–AI co-agency and delegation dynamics are insufficiently theorized. | Investigating GenAI as a "strategic collaborator" in ideation and decision-making processes can provide insights into evolving human–machine entrepreneurial strategy and ethical boundaries. | Sociotechnical systems theory; Cognitive science frameworks; Human–AI interaction models |
|---|---|---|---|---|
| **2 Generative AI-Enhanced Education & Learning Systems** | GenAI and LLM technologies are reshaping entrepreneurship education by transforming pedagogical practices, skill acquisition, and institutional frameworks. | A lack of longitudinal research limits understanding of GenAI's and LLMs' impact on learning and entrepreneurial capability development. | Future research should investigate the long-term pedagogical outcomes of GenAI and LLM integration in entrepreneurship curricula, particularly in fostering higher-order cognitive skills. | Longitudinal studies; Educational impact frameworks; Constructivist learning theory |
| | | Limited attention is given to the development of critical GenAI and LLM literacy and ethical usage among students. | Studies should explore instructional strategies that promote critical engagement with GenAI and LLM outputs, reinforcing reflective thinking and ethical discernment. | Digital literacy theory; Ethics in education; Responsible AI frameworks |
| | | There is a risk that overreliance on GenAI and LLM may inhibit original thinking and creativity for entrepreneurs. | Research should assess pedagogical designs that balance GenAI- and LLM-assisted learning with the cultivation of innovation, ambiguity tolerance, and creative problem-solving. | Creativity studies; Blended learning models; Entrepreneurial mindset development |
| | | Existing pedagogical models are not fully adapted to GenAI-augmented learning environments. | Scholars should develop new or revised pedagogical theories tailored to GenAI-driven learning, addressing assessment, learner–AI interaction, and co-creativity. | Instructional design theory; Human-AI collaboration frameworks |
| | | Institutional factors affecting GenAI integration remain underexplored. | Future work may examine how leadership, governance structures, and organizational culture mediate GenAI adoption in higher education settings. | Institutional theory; Change management in education |
| | | Prompt engineering as a core educational competency lacks empirical investigation. | Studies should assess pedagogical approaches to teaching prompt engineering and its impact on student engagement and learning efficacy. | Skill acquisition theory; Human–AI interaction models |
| | | Current GenAI and LLM tools exhibit limitations in semantic reasoning and domain-specific tasks. | Research should explore hybrid instructional strategies that integrate AI capabilities with | Hybrid learning models; AI affordance frameworks |
| | | | educator-guided mentorship to overcome these limitations. | |
| | | Insufficient methodological standards exist for documenting GenAI and LLM use in academic settings. | Scholars should develop clear reporting and documentation guidelines to ensure transparency and reproducibility in GenAI- and LLM-supported research and teaching. | Research ethics; Methodological standardization literature |
| | | Most research is confined to academic institutions, neglecting practical and informal entrepreneurial learning contexts. | Future research should investigate the application and outcomes of GenAI- and LLM-enhanced education in startup ecosystems, incubators, and accelerators. | Action research; Practice-based learning theories |
| | | Cross-cultural variability in GenAI's and LLM's educational impact is largely unaddressed. | Comparative studies should explore how diverse socio-economic, institutional, and cultural contexts mediate the integration and effectiveness of GenAI and LLM tools. | Cross-cultural education theory; Contextualized learning models |
| 3 Sustainable Innovation & Strategic AI Impact | GenAI enables new forms of sustainable innovation, human–AI collaboration, and ESG-aligned strategic entrepreneurship across diverse organizational settings. | Insufficient theorization of long-term human–AI collaborative processes in sustainable innovation. | Research should examine how human–AI collaboration evolves over time, and may be used to to support goals related to sustainability | Socio-technical systems theory; Human–AI teaming models |
| | | Context-specific variables (e.g., sector, region, firm size) are often underexplored. | Future studies should assess how contextual factors shape strategic GenAI adoption and value creation. | Contingency theory; Comparative industry studies |
| | | Inclusive access to GenAI for marginalized entrepreneurs is under-theorized. | Research should explore strategies to increase GenAI accessibility among underrepresented groups, including those with limited digital literacy or infrastructure. | Inclusive innovation theory; Digital equity frameworks |
| | | Legacy firms and SMEs face unique challenges in GenAI adoption, yet remain understudied. | Investigate the processes by which traditional firms integrate GenAI while maintaining trust, cohesion, and cultural alignment. | Organizational change theory; SME innovation models |
| | | Ethical implications of GenAI-driven efficiency gains are insufficiently analyzed. | Scholars should develop frameworks aligning GenAI applications with ESG metrics and responsible innovation standards. | Ethical AI frameworks; ESG integration models |

| | | Convergence between GenAI and other technologies (e.g., IoT, Web 3.0) lacks empirical validation. | Future research may assess how such technological synergies foster decentralized and regional innovation ecosystems. | Innovation ecosystem theory; Platform convergence models |
|---|---|---|---|---|
| | | Talent development and institutional entrepreneurship are emerging yet underexplored dimensions of GenAI's strategic role. | Studies should explore how GenAI influences human capital formation, learning mechanisms, and organizational renewal. | Dynamic capabilities theory; Institutional entrepreneurship |
| | | Policy and governance issues have not kept pace with GenAI's entrepreneurial impact. | Future research should investigate regulatory design, IP protection, and innovation governance in GenAI-centric entrepreneurship. | Policy analysis frameworks; Innovation law and governance models |
| **4 Business Models & Market Trends** | GenAI and LLMs are transforming entrepreneurial business models, startup ecosystems, and investor interactions through automation, analytics, and platform logic. | Existing studies focus on early-stage adaptation, with little understanding of long-term business model transformation. | Longitudinal research should explore the role GenAI and LLMs as external enabler for entrepreneurship and their scope, mechanisms and roles. | External Enablers for Entrepreneurship |
| | | | Evolutionary impact of GenAI and LLMs on entrepreneurial business models across varying levels of environmental turbulence. | Business model innovation frameworks; Longitudinal process studies |
| | | Socio-technical organizational dynamics are under-investigated. | Future studies should examine how structural and cultural configurations (e.g., openness) facilitate or inhibit GenAI adoption across firms and industries. | Organizational behavior theory; Comparative case methods |
| | | Investor roles in GenAI venture dynamics lack empirical grounding. | Research should assess how perceptions of GenAI maturity affect investment decisions and explore alternative finance models like DeFi and crowdfunding. | Entrepreneurial finance theory; Behavioral economics |
| | | Startup support systems and GenAI-enabled mentorship remain understudied. | Examine how virtual advisory systems and AI-augmented mentoring influence entrepreneurial learning and performance. | Entrepreneurial ecosystem theory; Digital mentorship frameworks |
| | | The communicative affordances of GenAI in marketing and branding are overlooked. | Future research should explore how GenAI supports narrative creation, identity work, and market differentiation. | Narrative identity theory; Brand communication models |
| | | Limited understanding exists around experiential | Empirical research should explore how iSMEs apply | Effectuation theory; Bricolage; Learning-by-doing frameworks |
| | | learning in international SMEs (iSMEs) using GenAI and LLMs. | effectuation, causation, and bricolage when integrating GenAI and LLMs. | |
| | | Emerging paradigms (e.g., meta-entrepreneurship, immersive ecosystems) are under-theorized. | Future studies should theorize GenAI-enabled hybrid business models that incorporate technologies like blockchain and the metaverse. | Theory-building; Socio-technical system integration frameworks |
| | | Entrepreneurship education's evolving role remains inadequately studied. | Investigate how GenAI tools facilitate interdisciplinary learning, sustainability orientation, and educational inclusion. | Inclusive education models; Interdisciplinary curriculum design |
| 5 Data-Driven Tech Trends in Entrepreneurship | Data-centric GenAI, LLM (e.g. ChatGPT) tools are redefining resource use, planning, strategic decision-making, and systemic innovation processes in entrepreneurship. | Contextual variability in GenAI and LLM adoption remains underexplored. | Comparative studies should investigate how regulatory environments, infrastructure quality, and institutional logics shape GenAI and LLM deployment and the possibility to realize their business and entrepreneurial potential | Institutional theory; National innovation systems, innovation studies and entrepreneurship |
| | | Multi-modal data integration in entrepreneurship is not well understood. | Future work should analyze how entrepreneurs fuse behavioral, transactional, and sensor data to support strategic agility. | Organizational capability theory; Data fusion analytics |
| | | AI as a strategic resource requires further theorization. | Research should examine the micro-foundations of AI resource orchestration and deployment for competitive advantage. | Resource-Based View (RBV); Micro-foundations of strategy |
| | | The role of infrastructure and policy in enabling or constraining AI adoption is under-analyzed. | Longitudinal and regional analyses should assess how smart infrastructure investments translate into entrepreneurial outputs. | Spatial analysis; Policy and infrastructure studies |
| | | Methodological innovation using AI in research lacks scrutiny. | Meta-methodological studies should evaluate the epistemic, ethical, and practical implications of AI-based research tools. | Research ethics; Methodological transparency frameworks |
| | | Equity and inclusion issues in GenAI access persist. | Research should investigate mechanisms (e.g., decentralized AI) for democratizing access to AI technologies in entrepreneurship. | Socio-technical inclusion models; Open-source governance |
| | | Systemic effects of GenAI and LLMs on innovation dynamics | Future studies should explore how GenAI and LLMs reshape competition, entry | Systems theory; Innovation governance; Competitive dynamics |

| | | are not well understood. | barriers, and monopolization trends in entrepreneurial ecosystems. |
|---|---|---|---|

Source: authors' own elaboration based on the research clusters.

By addressing the identified areas across the five clusters, future research can capture how GenAI and LLM can be used and developed within entrepreneurial contexts. This includes promoting a pedagogically informed integration of GenAI into entrepreneurship education, particularly for the next generation of founders. It is also essential for future research to examine policy-related issues, including regulation, intellectual property, and responsible innovation, to ensure that regulatory frameworks allow for experimentation and the realization of the business and innovation potential of GenAI, while recognizing issues and goals related to the ethics and inclusiveness. As data-driven technologies such as GenAI and LLM become increasingly embedded within entrepreneurial ecosystems, it is essential for future research to explore their complex interplay with institutional, technological, and socio-economic environments. Collectively, these research directions offer a foundation for developing a more inclusive, adaptive, and future-oriented understanding of entrepreneurial potential of GenAI.

A general remark on the current state of research on GenAI and LLMs in the context of entrepreneurship is that there is a need for more "macro-level" analyses. With this we mean that a) there is a need to better understand how GenAI and LLMs influence the preconditions for entrepreneurship and business dynamics as well as b) how regulatory frameworks may need to be adapted to pave the way for the full realization of the business and innovation potential for GenAI and LLMs.

Regarding the former, GenAI and LLMs can improve the efficiency and innovation potential of established firms as well as new innovative entrants and there is a need to better understand what factors that drive entrepreneurship in the form of innovative entrants. We believe that research on GenAI and LLMs as external enablers for entrepreneurship that assess their scope, mechanisms and potential roles is one relevant avenue for future research (Kimjeon and Davidsson 2022).

Regarding the latter, as pointed out before, it is well established that there are several calls for regulation of the use of GenAI and LLMs in business and society at large to adress issues of ethics and lower risks of various types of misuse of the technologies. However, there is a lack of research that address the potential "goal conflicts" involved in work on regulatory frameworks. How can we develop regulatory frameworks that secure ethical concerns and prevent various types of misuse but at the same time facilitate business experimentation, innovation and further technology development that allow for economy-wide learning?

# 6. Conclusion

To the best of our knowledge, this is the first systematic literature review of GenAI and LLMs in research on entrepreneurship. Using a dataset of 83 peer-reviewed academic publications sourced from Scopus and Web of Science, and employing advanced natural language processing and unsupervised machine learning techniques, the study identifies five major thematic clusters that collectively reflect the current state and future trajectory of this interdisciplinary field: (1) Digital Transformation and Behavioural Models, (2) GenAI-Enhanced Education and Learning Systems, (3) Sustainable Innovation and Strategic AI Impact, (4) Business Models and Market Trends, and (5) Data-Driven Technological Trends in Entrepreneurship.

The findings highlight that existing research emphasizes the role GenAI plays across entrepreneurial processes, from enhancing cognitive mechanisms and opportunity recognition to enabling data-driven decision-making, developing innovative business models, and redefining entrepreneurial education. GenAI is not just a new "technological tool", but a technology that has a potentially significant impact on the socio-technical environment and interacts with institutional contexts and pedagogical frameworks. The behavioural and educational implications emphasize the need to approach GenAI adoption through interdisciplinary lense that account for both technological affordances and human-centred dynamics.

Moreover, the paper identifies several ethical concerns including data privacy, algorithmic bias, intellectual property, misinformation and others that must be addressed as GenAI becomes increasingly embedded in entrepreneurial ecosystems. These issues relate to current discussions about development of regulatory frameworks, transparency mechanisms, and ethically-aligned design principles regarding the integration of GenAI technologies, often motivated with references to responsible innovation and inclusive value creation. We also pinpoint the need for more "macro-level" research on GenAI and LLMs as external enablers for entrepreneurship and effective regulatory frameworks that allow for business experimentation, innovation and further technology development.